\documentclass[algo2e]{ifacconf}

\usepackage{graphicx}      
\usepackage{natbib}        
\usepackage{url}

\usepackage{amsmath}
\usepackage{amsfonts}
\usepackage{amssymb}
\usepackage[nolist]{acronym}

\usepackage{listings}
\usepackage{multicol}
\usepackage{subcaption}
\usepackage{enumerate}
\usepackage{graphicx}  
\usepackage{xcolor}
\usepackage{float}

\usepackage{siunitx}
\usepackage{xspace}

\usepackage[ruled]{algorithm2e}
\usepackage[utf8]{inputenc}

\usepackage{color}
\usepackage{mathtools}

\usepackage{multirow}

\usepackage{eurosym} 

\usepackage{nicefrac}

\usepackage{smartref}

\newcommand{\todo}[1]{}
\newcommand{\figref}[1]{Fig.~\ref{#1}}

\newcommand{\qm}[1]{``{#1}''}

\newcommand{\Eth}{\ensuremath{E_\mathrm{th}}\xspace}
\newcommand{\Ebd}{\ensuremath{E_\mathrm{bd}}\xspace}
\newcommand{\hl}{\ensuremath{h_\mathrm{l}}\xspace}

\newcommand{\Jd}{\ensuremath{J^\mathrm{d}}\xspace}
\newcommand{\tf}{\ensuremath{t_\mathrm{f}}\xspace}

\newcommand{\Jps}{\ensuremath{J_\mathrm{ps}}\xspace}
\newcommand{\iw}{\ensuremath{i_\mathrm{w}}\xspace}
\newcommand{\tbd}{\ensuremath{t_\mathrm{bd}}\xspace}
\newcommand{\nw}{\ensuremath{n_\mathrm{w}}\xspace}

\newcommand{\xvek}{\ensuremath{\xi}}
\newcommand{\ign}[1]{}

\newtheorem{problem}{Problem}

\DeclareMathOperator*{\minimize}{minimize}
\DeclareMathOperator*{\argmax}{arg\, max}
\newcommand{\st}{\text{subject to}}

\graphicspath{{images/}}
\begin{document}
	\begin{frontmatter}
		
		\title{Multi-Objective Model-Predictive Control for Dielectric Elastomer Wave Harvesters} 
		
		\author[First]{Matthias K. Hoffmann} 
		\author[First]{Lennart Heib} 
		\author[Second]{Gianluca Rizzello}
		\author[Third]{Giacomo Moretti}
		\author[First]{Kathrin Flaßkamp} 
		
		\address[First]{Systems Modelling and Simulation, \linebreak Saarland University, Saarbrücken, Germany \linebreak (e-mail:~\{{matthias.hoffmann, kathrin.flasskamp\}@uni-saarland.de, lennartheib@gmail.com})}
		\address[Second]{Adaptive Polymer Systems, \linebreak Saarland University, Saarbrücken, Germany.	}
		\address[Third]{Department of Industrial Engineering, Università di Trento, Italy.}
		
		\begin{abstract}                
		This contribution deals with multi-objective \ac{mpc} of a \ac{wec} device concept, which can harvest energy from sea waves using a \ac{deg} power take-off system. 
		We aim to maximise the extracted energy through control while minimising the accumulated damage to the DEG.
        With reference to system operation in stochastic waves, we first generate ground truth solutions by solving an optimal control problem, and we analyse the MPC performance to determine a prediction horizon that trades off accuracy and efficiency for computation.\
		Fixed weights in the MPC scheme can produce unpredictable costs for variable sea condition, meaning the average rate of cost accumulation can vary vastly. To steer this cost growth, we propose a heuristic to adapt the algorithm by changing the weighting of the cost functions using for fulfilling the long-time goal of accumulating a small enough damage in a fixed time.\ 
		A simulated case-study is presented in order to evaluate the performance of the proposed MPC framework and the weight-adaptation algorithm.\
		The proposed heuristic proves to be able to limit the amount of accumulated damage while remaining close to (or even improving) the energy yield obtained with a comparable fixed-weight MPC. 
		\end{abstract}
		
		
	\end{frontmatter}
	\begin{acronym}
	\acro{deg}[DEG]{dielectric elastomer generator}
	\acro{moo}[MOO]{multi-objective optimization}
	\acro{ocp}[OCP]{optimal control problem}
	\acro{moocp}[MOOCP]{multi-objective optimal control problem}
	\acro{wec}[WEC]{wave energy converter}
	\acro{mpc}[MPC]{model-predictive control}
	\acro{pop}[POP]{Pareto optimal point}
	\acro{nlp}[NLP]{non-linear program}
	\acro{foh}[FOH]{first-order hold}
\end{acronym}
	
	\section{Introduction}
Ocean wave energy is a highly abundant and dense form of renewable energy. 
Although many different concepts of \acfp{wec} were studied in the past, their high technological complexity and deployment costs have hindered these technologies from being used in the field (\cite{Pecher2017}).\ 
One promising approach to overcome these barriers is the use of \acfp{deg}, i.e. lightweight polymeric generators based on low-cost raw materials, which allow direct conversion of mechanical energy into electrical energy based on a variable-capacitance principle (\cite{Moretti2020}).\\
In a previous publication, we derived a model-based optimal control for a \ac{deg}-\ac{wec} subject to sinusoidal regular waves (see \cite{Hoffmann2022moocp_wcdeg}).
We showed that multi-objective optimal control can, under consideration of a non-linear model, disclose technically relevant trade-offs between the electrical damage accumulated by the \ac{deg} over time and the extracted energy, allowing for a reduction of damage by more than 50~\% while only losing 1~\% of energy compared to a control that only aims at maximising the power.\\ 
However, these results were based on the assumption that the wave motion is periodic and can be predicted far into the future.
In reality, ocean waves are irregular and therefore unpredictable for long time-horizons (\cite{Coe2018}).\ This makes the use of open-loop optimal control difficult, as errors in the prediction of the wave excitation result in suboptimal control signals.\ Under the assumption that a correct prediction shortly into the future is possible, \acf{mpc} can be used to generate a suitable control signal during operation, having better adaptability to environmental changes than optimal control, while preserving the same cost functions and constraints (\cite{Faedo2017}).\
The \ac{mpc} algorithm computes control inputs by solving an \ac{ocp} with shorter prediction horizon and updating the input estimation in-line, while controls are executed.\\
In this work, we apply for the first time \ac{mpc} to \ac{deg}-\acp{wec} under the influence of stochastic waves.
One key question our paper aims to answer is how long the \ac{mpc} prediction horizon needs to be, so that the resulting control does not deviate from an optimal solution.\ 
For that, we evaluate the deviation of the \ac{mpc} solution with respect to a ground truth \ac{ocp} input signal to find that reasonable prediction horizon lengths have to include multiple wave peaks.\ 
Additionally, if long-time goals are to be achieved, it is needed that the controller explicitly accounts for the high variability in the system operating conditions.\
In a real-world application it would be beneficial to track the accumulation of damage on the \ac{deg}, so that a prediction of the time of failure is possible. This might, among other, allow 
actively adjusting the controller parameters as the system approaches breakdown, similar to ideas presented by \cite{Requate2022}. With this objective in mind, we designed a simple heuristic switching scheme that can adapt the \acp{mpc}. 
This adaptation is achieved using multi-objective \ac{mpc} with a simple heuristic for changing the weights in the multi-objective optimal control scheme.
We show that even a rudimentary switching scheme is effective in limiting the damage accumulation below an arbitrary threshold.

	\section{Model and Problem Statement}

\begin{figure}[htb]
	\centering
	\fontsize{9}{0}\selectfont
	\def\svgwidth{0.5\textwidth}
	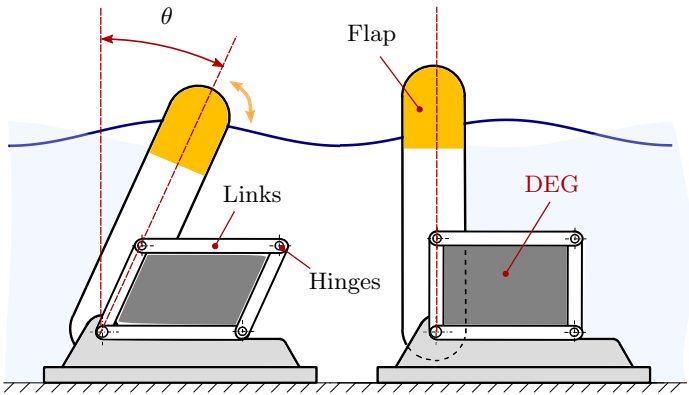
	\caption{Wave surge converter: a flap hinged on the sea floor is tilted by the wave motion. It is displayed in a generic (left) and the vertical equilibrium position (right).}
\label{fig:flap}
\end{figure}

\subsection{System description}
In this work, we design an \ac{mpc} scheme for the wave surge converter (see, e.g. \cite{Whittaker2012}), displayed in \figref{fig:flap}.\
The device is hinged to the sea bed, so that incoming waves excite a pitch motion.\ 
This deforms a \ac{deg}, which consists in a stack of electrode-covered polymeric dielectric membranes, whose perimeters is rigidly connected to a parallelogram mechanism (\cite{Moretti2014}).\ 
When no voltage is applied, the \ac{deg} generate an elastic torque that pushes the flap towards the vertical equilibrium position $\theta=0$.\ 
Applying a voltage to the \ac{deg} adds an electrostatically-induced torque in the same direction as the elastic torque, making the system stiffer. The elastic torque can be considered negligible compared to the electrostatic torque (\cite{Moretti2014}).\ 

The \ac{deg} functions as a variable capacitor, which can be controlled so as to  generate electrical energy at the expense of the input mechanical work generated by the sea wave moving the flap. E.g., applying low or no voltage during the phases in which the capacitance increases ($\theta\dot{\theta}<0$) and a high voltage in the phases when the the capacitance decreases ($\theta\dot{\theta}>0$) causes current to flow out of the \ac{deg} and electrical power to be delivered to the power electronics.\
As we showed in our previous work \cite{Hoffmann2022moocp_wcdeg}, controlling the input voltage to maximise the energy extracted from the system results in large large electric fields, damaging the \ac{deg}-material over time and leading to system failure after damage reaches a certain threshold (\cite{Chen2019}).\
For that reason, we also took the damage into consideration as a second optimization objective and, in turn, consistently limit the electric field in the DEG.

\subsection{Model}
For small oscillation angles, the dynamics of the wave surge takes the following form (\cite{Hoffmann2022moocp_wcdeg}):
\begin{equation}\label{eq:dynamics}
	\begin{split}
		\begin{bmatrix}
			\dot{\theta} \\
			\dot{\delta}  \\
			\dot{z}
		\end{bmatrix}
		& = \begin{bmatrix}
			0 & 1 & 0^{1\times n} \\
			-I_h^{-1}K_\mathrm{h} & -I_\mathrm{h}^{-1}B_\mathrm{h} & -I_\mathrm{h}^{-1}C_\mathrm{r} \\[3pt]
			0^{n\times 1} & B_\mathrm{r} & A_\mathrm{r}
		\end{bmatrix}    \begin{bmatrix}
			{\theta} \\
			{\delta}  \\
			{z}
		\end{bmatrix} \\ &  + \begin{bmatrix}
			0 \\
			I_\mathrm{h}^{-1}  \\
			0^{n\times 1}
		\end{bmatrix} \left(d - C_0\theta u \right)\\
		\theta(0) &= \theta_0,\ \delta(0) = \delta_0,\ z(0) = z_0.
	\end{split}
\end{equation}
where $\theta$ and $\delta = \dot{\theta}$ describe the flap's angular position and velocity; $z \in \mathbb{R}^n$ is an $n$-dimensional state vector describing the dynamics of the radiated waves'  force (\cite{Yu1995}); $K_h$ and $B_h$ represent the hydrodynamic stiffness (due to buoyancy) and damping of the flap; $A_r,\ B_r,\ C_r$ are matrices modelling the radiated waves dynamics; $d$ is a time-varying torque the waves exert on the flap;  input $u$ physically represents the voltage $v$ applied on the \ac{deg} squared. 
For a derivation of this model refer to our previous work (\cite{Hoffmann2022moocp_wcdeg}).\
In the following, the system's state is denoted by $x = [\theta, \delta, z^\intercal]^\intercal$. In operating conditions, the input $u$ should meet the following hard constraints:
\begin{equation}\label{eq:constr}
	0 \le u \le (\Ebd \hl)^2,
\end{equation}
where the first inequality of $0 \le u$ owes to the definition of $u$ ($u=v^2$), whereas the second constraint demands that the electric field is lower than a threshold breakdown value, $\Ebd$, that would cause static failure of the \ac{deg}.

\subsection{Cost functions}
Our aim is the simultaneous minimisation of accumulated damage and maximisation of extracted energy, which we will employ in a \ac{moocp} setting.\ 
The  energy cost function is defined as the  energy generated over a time $t_f$ with changed sign:
\begin{align}\label{eq:J1}
		& J_1(x(t), u(t), \tf) = \Psi(x(t), u(t), \tf)-\Psi(x(t), u(t), 0)\nonumber\\
		& \hspace{1,2cm} + \int_0^{\tf} \left(B_\mathrm{h}\delta(t)^2 +z(t)^\intercal S_\mathrm{r} z(t) +\dfrac{u(t)}{R_0} - d \delta(t)\right) \text{d}t \nonumber\\
		& \text{with}\ \Psi(x(t), u(t), \tau) = \dfrac{1}{2}I_\mathrm{h}\delta(\tau)^2+\dfrac{1}{2}K_\mathrm{h}{\theta(\tau)}^2 \nonumber\\
		&\hspace{1,2cm} + \dfrac{1}{2}z(\tau)^\intercal Q_\mathrm{r} z(\tau) + \dfrac{1}{2} C_0\left(1 - \theta(\tau)^2 \right) u(\tau),
\end{align}
with the storage function $\Psi$ including kinematic,  electrostatic, and hydrostatic energy contributions.\ 
Dissipations due to viscous, hydrodynamic and electrical losses, and the power input by the incident wave are considered via the integral term in \eqref{eq:J1}.\ 
Under the assumptions that the electric field is the main source of damage for the DEG (\cite{Chen2019}) and that damage only starts accumulating if the  electric field exceeds a threshold value $\Eth$ (\cite{Dissado1992}), the damage cost function can be formulated as
\begin{align}\label{eq:J2}
	&J_2(x(t), u(t), \tf) = \nonumber\\
	&\qquad\alpha \int_0^{\tf} \left(\max \{ u(t)-\Eth^2 {\hl}^2,0 \} \right) \text{d}t, 
\end{align}
with a normalisation factor $\alpha$ rendering $J_2$ dimensionless.\ 

Equations \eqref{eq:dynamics}, \eqref{eq:constr}, \eqref{eq:J1}, and \eqref{eq:J2} define the \ac{moocp}

\begin{problem}\label{prob:ocp}
	\begin{align}\label{eq:problem}
	&\minimize_{u} \ \ \, (J_1, J_2) \nonumber\\
	\ &\st \ \text{dynamics}\ \eqref{eq:dynamics} \nonumber\\
	&0 \le  u \le (\Ebd \hl)^2.
\end{align}
\end{problem}
that is solved inside the \ac{mpc} framework.
	\section{Methods}

\subsection{Background on MPC}
\begin{figure}[htb]
	\centering
	\fontsize{8}{0}\selectfont
	\def\svgwidth{0.49\textwidth}
	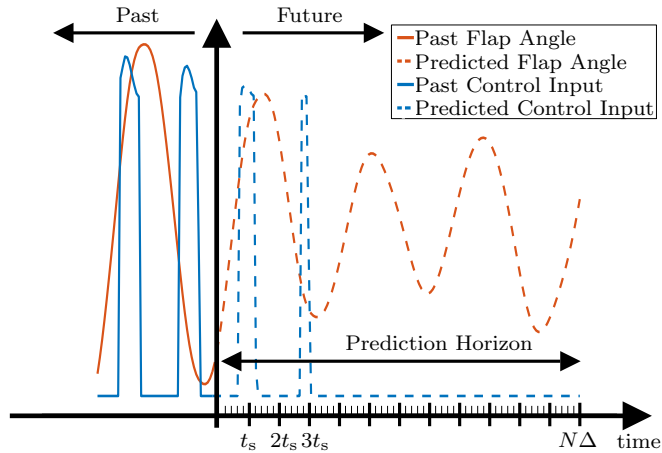
	\caption{The MPC working principle on the example of the DEG-WEC.\ The solid line shows the state and input history, while the dashed line displays the prediction that will be applied to the system partially.}
	\label{fig:mpc_explanation}
\end{figure}

\ac{mpc} arose from optimal control as an answer on how to \qm{close the loop} in model-based open-loop optimal control (\cite{rawlings2017mpc}).\ 
In optimal control, a system's behaviour is predicted for a time called the prediction horizon \tf into the future, while optimising the inputs to the system in such a way that a cost function is minimised.\
The workflow of \ac{mpc} consists in repeatedly measuring the system's state (or estimating it with an observer), solving an \ac{ocp}, and feeding the first portion of the calculated inputs to the plant.\
\figref{fig:mpc_explanation} qualitatively shows the state and input history and the predicted state and input that will partially be applied to the system.\

\subsection{Discretisation and simplification of the \ac{ocp}}
In order to solve Problem~\ref{prob:ocp}, we employ direct methods for optimal control, i.e. the OCP is transcribed into a non-linear program which is then solved by appropriate methods (see \cite{Gerdts2011OC}).
 Using gradient-based methods, the discretised optimal control signal is calculated.\ 
The integral terms inside the cost functions have to be discretised as well.\ 
We do so by adding the integrand to the dynamics
\begin{align*}
	\dot{\Upsilon}_1 &= B_h\delta^2 +z^\intercal S_r z +\dfrac{u}{R_0} - d \delta \nonumber\\
	\dot{\Upsilon}_2 &= \max \{ u-E_{th}^2 h_l^2,0 \},
\end{align*}\ 
with
\begin{align*}
	\Upsilon_1(0) = \Upsilon_2(0) = 0.
\end{align*}
The extended state reads
$\xvek = \begin{bmatrix}
	{\theta} &
	{\delta} & 
	{z{^\intercal}} &
	{\Upsilon_1} &
	{\Upsilon_2}
\end{bmatrix}^\intercal$, with the initial value $\xvek_0 = \begin{bmatrix}
{\theta_0} &
{\delta_0} & 
{z_0{^\intercal}} &
{0} &
{0}
\end{bmatrix}^\intercal$

For a time step denoted by $\Delta$ and some final time $t_f = N\Delta$, we now introduce a time discretization $\{ k\Delta \}_{k=0}^N = \{0, \Delta, 2 \Delta,\dots , N \Delta\}$.\ In the following, the discretised values corresponding to their continuous counterparts are marked by square brackets, e.g. $\xvek[k]$ denotes the extended state $k$ time steps into the future.\ 
The current state of the system is advanced by one step into the future using the classical Runge-Kutta-Method of 4-th order (RK4), denoted by $F_\mathrm{RK4}(\xvek[k], u[k], u[k+1], d[k], d[k+1])$.\ 
Consecutive values for the input and wave excitation are used to model \ac{foh} behaviour.\ 
The dynamics can then be expressed with the equality constraints
\begin{align*}
	\xvek[k+1] = F_\mathrm{RK4}(\xvek[k], u[k], u[k+1], d[k], d[k+1]) \\
	\forall k\in\left[0,N-2]\right]\\
	\xvek[0] = \xvek_0.
\end{align*} 
The cost functions are then $\tilde{J}_1 = \Upsilon_1[N-1]$ and $\tilde{J}_2 = \Upsilon_2[N-1]$, so that the \ac{moocp} is
\begin{problem}\label{pb:problem_disc}
	\begin{align}\label{eq:problem2}
		\minimize_{u[1],\ldots,u[N]} \ &w_1 \tilde{J}_1 + w_2 \tilde{J}_2 \nonumber\\
		\ \st \ & \xvek[k+1] = F_\mathrm{RK4}(\xvek[k], u[k], u[k+1],\dots\nonumber\\ &d[k], d[k+1])  \forall \ k \in [0, N-2], \nonumber\\
		&0 \le  u[k] \le (E_{bd} h_l)^2, \ \forall \ k \in [0, N-1]\nonumber\\
		& \xvek[0] = \xvek_0 \nonumber\\
		& u[0] = u_0.
	\end{align}
\end{problem}
Within the MPC, Problem~2 is solved repeatedly on shifted time intervals and updated initial states.\ 
As depicted in \figref{fig:mpc_explanation}, $N\Delta$ becomes the prediction horizon and, let us assume, for the sampling time we have $t_s = r \Delta, r\in \mathbb{N}_{>0}$.\
That is, in each sampling step, the controls u[0],...u[r] are applied to the plant while the rest is discarded.\
Then, $\xi_0$ is set to the current state, $u_0$ is set to $u[r]$, and the MOOCP is solved again.
The resulting control signal is denoted by $u_\mathrm{MPC}$.
The values of $\Psi$ are omitted in the formulation.\ 
Since $\Psi$ for $\tau=0$ is a constant in the \ac{foh} formulation, it does not change the optimization problem.\ 
Regarding $\Psi$ at $t_f$, since $I_\mathrm{h}, K_\mathrm{h}$ are orders of magnitude larger than $C_0$, their terms dominate the value of $\Psi$.\ 
The quadratic cost terms in $\Psi$ push the solution to the equilibrium position $\theta = 0$ at the end of the prediction horizon, an effect unwanted in continuous operation, so it is omitted from the energy cost function.
\textit{Notation:} The subindex $\mathrm{MPC}$ marks applied inputs and resulting states of the actual system.

\subsection{Generation of the wave excitation profiles}
The stochastic wave is modelled as a superposition of a number of $n_\mathrm{f}$ sine waves.\ The amplitude of the different harmonic components has a distribution described by the so-called Bretschneider spectrum. The Bretschneider spectrum 
\begin{align*}
	& S_B(\omega)=A_\mathrm{B} \omega^{-5} \exp \left(-B_\mathrm{B} \omega^{-4}\right)
\end{align*}
describes an average sea state when wave elevation profile measurements are not available, where $\omega$ represents an angular frequency. 
The wave excitation torque
\begin{align*}
	d(t)=\sum_{i = 1}^{n_\mathrm{f}} \Gamma_\mathrm{F}(\omega_i) A_i(\omega_i) \sin \left(\omega_i t+\phi_i(t)\right).
\end{align*}
can be calculated from the spectrum, where $\Gamma_\mathrm{F}(\omega)$  is a frequency-dependent excitation coefficient (depending on the hydrodynamics). The coefficients $A_i$ represent the amplitudes of the different harmonic components in the wave profile, given by:
\begin{align*}
	A_i=\sqrt{2 S_\mathrm{B}(\omega_i) \Delta \omega_i},
\end{align*}
where $\Delta \omega_i$ are frequency increments, $\phi_i(t)$ are random phase-shifts, and the wave frequencies $\omega_i = \omega_0 + i\Delta\omega \ \forall \ i \in [0, n_\mathrm{f}] $ are linearly increasing.
With that, let $$\omega_\mathrm{f} = \left\{\omega_i \ :  \ i = \argmax\limits_i A_i, i \in [0, n_\mathrm{f}]\right\}$$ be the dominant frequency of the wave.
The parameters $A_\mathrm{B}$ and $B_\mathrm{B}$ can be modified to yield a wave with the desired overall significant wave height, $\omega_\mathrm{f}$ and frequency profile. For this paper, no specific sea state is emulated.\\
Here, $\phi_i(t)$ are set to change slowly over time so as to prevent the onset of any periodicity in the generated excitation. \ 
For the purpose of exemplification, in this paper spectral parameters $A_\mathrm{B}$ and $B_\mathrm{B}$ 
are simply chosen in such a way as to keep the resulting trajectories of $\theta$ within the model validity bounds, rather than with the aim of representing location-specific sea states.\


\subsection{Adaptive weight selection}
\label{sec:weight_control}
\begin{figure}[htb]
	\centering
	\fontsize{8}{0}\selectfont
	\def\svgwidth{0.4\textwidth}
	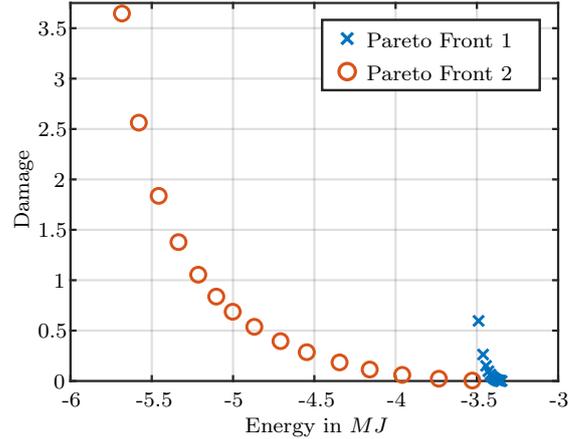
	\caption{Pareto fronts relative to two different realisations of a sea state (with same wave spectrum).}
	\label{fig:PF}
\end{figure}

When applying \ac{mpc}, we do not know exactly how the controlled system will perform cost-wise.\ 
Partially responsible for that is the change of what a set of weights means for different sea states.\ 
\figref{fig:PF} shows two Pareto fronts (relative to two realisations of a same wave spectrum) for a prediction horizon of \SI{60}{\second} for 15 evenly distributed weights between 0.05 and 0.95.\
Negative energy corresponds to extracted energy.  

In the case of the WEC-DEG, when driving the system with a fixed weighting, different sea states will result in a different damage accumulation over time.\ 
When the accumulated damage surpasses a certain threshold,  the \ac{deg}  breaks down and needs to be replaced. In a wave farm with multiple \acp{deg}, it might be desirable to render the average expected lifetime of  a certain device as close as possible to that of the rest of the farm, in a way that multiple units can be replaced together at once, hence minimising the operational costs. A strategy to extend the (average) expected lifetime of a unit up to a target time \tbd
is by changing the weighting of the damage cost function in a way that the accumulated damage cost at  time \tbd  does not exceed a fixed value $\Jd$ (corresponding to a rupture threshold).\ 
Let us consider a fixed set $(w_1, w_2)$ of \nw weight combinations with increasing values of $w_2$ (the weight of damage cost $J_2$), decreasing values of $w_1$, and an initial weight index $\iw \in [1, \nw]$.\ 
A way of approximating the future damage accumulation and by that estimating if the damage goal is achievable with the current weighting is by evaluating the \ac{mpc} performance over $N_p$ time steps into the past.\ 
The average rate of damage accumulation \Jps is estimated and the damage at the break-down time is predicted by assuming that the average damage accumulation trend continues as in the past $N_p$ steps.\
If the predicted damage exceeds \Jd, \iw is decreased by 1.\ 
Otherwise, if the predicted damage falls below $c_\mathrm{d}\Jd$ with $c_\mathrm{d}\in\left[0, 1\right]$, \iw is increased by 1.
This is done every $N_p$ steps (provided that the \ac{deg} was actuated with non-zero input during that time).\
We hereby show that  this very simple heuristic is effective in providing margins to extend the \ac{deg} lifetime, motivating further research on more elaborate adaptation algorithms.\ 

\begin{figure}[htb]
	\centering
	\fontsize{9}{0}\selectfont
	\def\svgwidth{0.48\textwidth}
	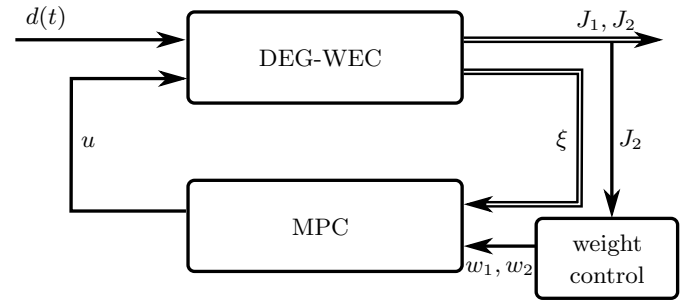
	\caption{Schematic of the weight controlled MPC. The weight controller keeps a history of past damage values, projects the estimated damage trend
 into the future and   selects a weighting to be used in the MPC accordingly.}
	\label{}
\end{figure}

%

	\section{Numerical Results}

We present numerical implementations of the proposed MPC framework in  MATLAB.\ The numerical data used for the analysis are the same as in \cite{Hoffmann2022moocp_wcdeg}. 
The optimisation problems were formulated and solved using the CasADi package by \cite{Andersson2019casadi} and the IPOPT solver by \cite{Waechter2006ipopt}.\
The stochastic waves were generated using a superposition of equally-spaced 50 harmonics with a base frequency of \SI{0.1}{\hertz} with $A_\mathrm{B} = 0.0032$ and $B_\mathrm{B} = 0.1054$. The nominal MPC case is assumed.\\
Examples of MATLAB implementations for the reference \ac{deg}-\ac{wec} system are available in our Github repository, with examples for fixed-weight and weight-controlled MPC\footnote{\url{https://github.com/MKHoffmann/IFAC_WC_2022_WaveHarvestingExample}}.

\subsection{Accuracy of model-predictive control}
\label{sec:accuracy}
\begin{figure}[htb]
	\centering
	\fontsize{8}{0}\selectfont
	\def\svgwidth{0.47\textwidth}
	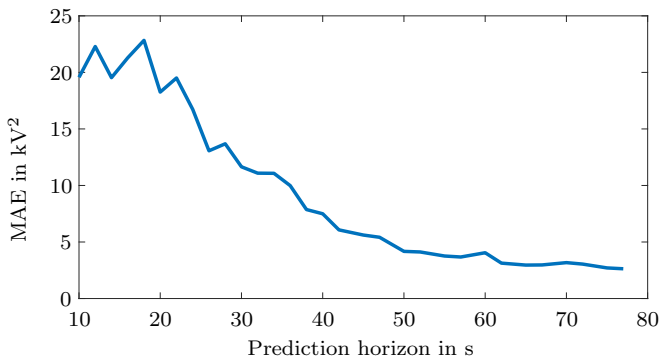
	\caption{Deviation of $u_\mathrm{MPC}$ from ground truth for different prediction horizon lengths. We take the mean error of simulations over a time window of \SI{320}{\second}.\ Ground truth is an OCP solution over the whole horizon.}
	\label{fig:mpc_error}
\end{figure}

Applying \ac{mpc} will, in general, generate different control trajectories compared to those obtained by solving 
an \ac{ocp} with very long prediction horizon.\ 
By shifting the prediction horizon at each time step, new information is provided, potentially leading to significantly different input sequences compared to previous solutions.\
\figref{fig:mpc_error} shows the mean absolute error (MAE), expressed as the $L_2$ norm of the difference between the ground truth \ac{ocp} solution over a prediction horizon of \SI{320}{\second} and the control signal obtained from MPC $u_\mathrm{MPC}$;\ 
$u_\mathrm{MPC}$ was calculated using different horizon lengths from 10 to \SI{77}{\second}.\
As expected, the error decreases for longer prediction horizons.\ 

\begin{figure}[htb]
	\centering
	\fontsize{8}{0}\selectfont
	\def\svgwidth{0.47\textwidth}
	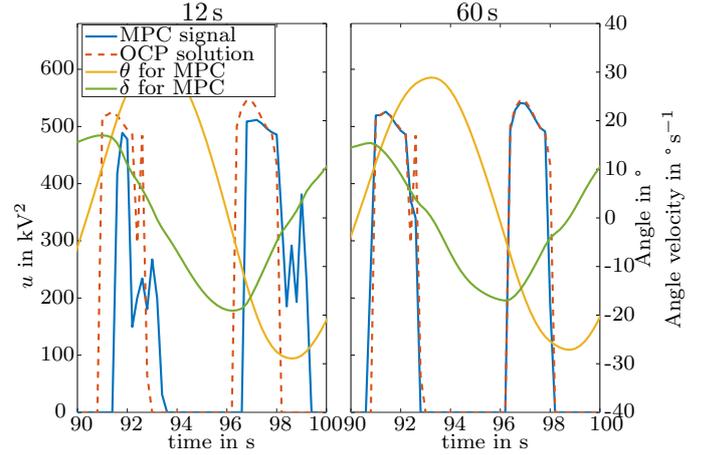
	\caption{Comparison of the MPC solutions for prediction horizons of \SI{12}{\second} and \SI{60}{\second} with the ground truth OCP solution. }
	\label{fig:mpc_comparison}
\end{figure}

In \figref{fig:mpc_comparison}, we show the control inputs and states for example solutions from different prediction horizons.
The ground truth optimal solution is approximated  poorly for a prediction horizon of \SI{12}{\second} (left) compared to an accurate tracking for \SI{60}{\second} (right).\ 
In all cases, the control has a bang-bang-like behaviour, with voltage being applied on the \ac{deg} only during certain time intervals. Short-horizon MPC solutions differ from the ground truth solution in terms of the turn-on and turn-off timings, which are correctly estimated when longer prediciton horizons are used.\ 
These trends are consistent with heuristic controllers proposed by \cite{Moretti2014}, in which a (piecewise constant) input is applied
 when $\theta\dot{\theta}\ge0$. Compared to such heuristic, the MPC leads to  complex  (non-piecewise-constant) voltage input waveforms.\
The similarity of the \SI{60}{\second}-MPC solution to the ground truth is also reflected by the extracted energy, which differs from the ground truth value by only 0.5 \%. 
In the following analyses, we will refer to an MPC horizon of \SI{60}{\second}.

\begin{figure*}[htb!]
	\centering
	\fontsize{8}{0}\selectfont
	\def\svgwidth{0.97\textwidth}
	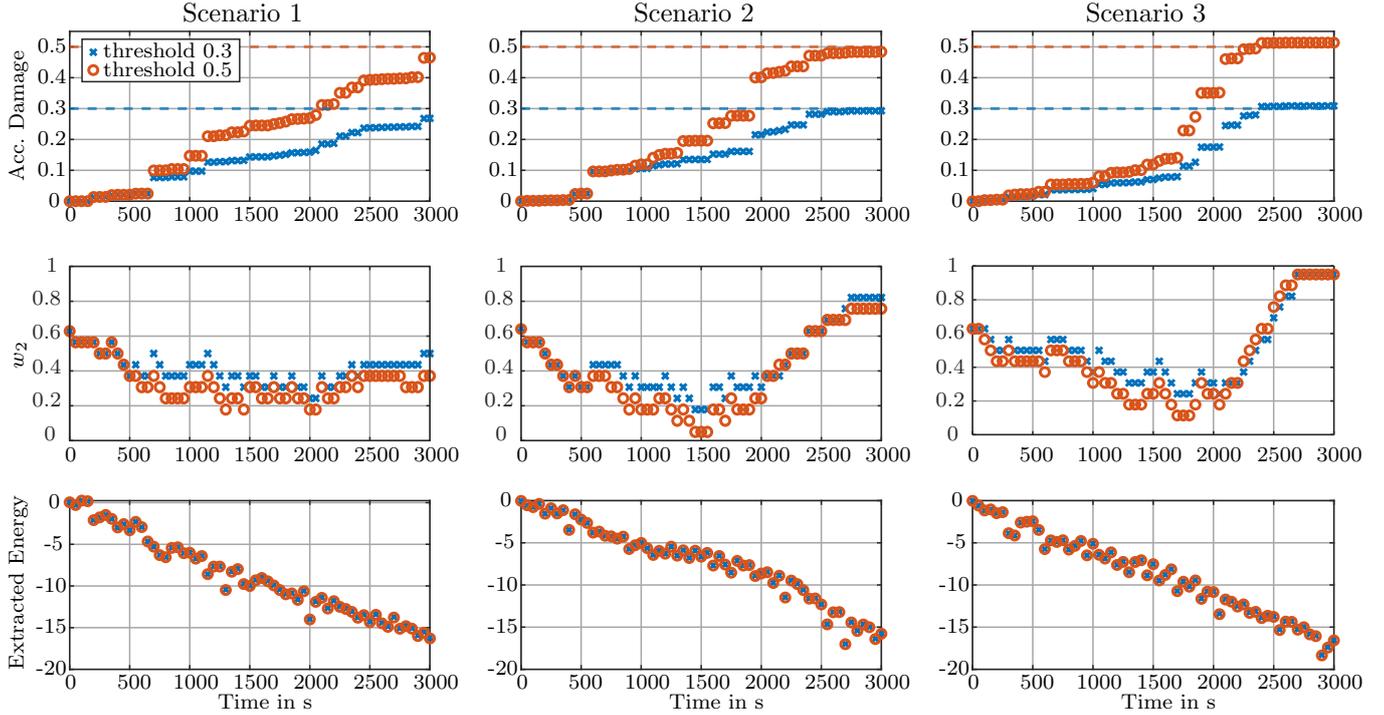
	\caption{Evaluation of the \ac{mpc} with weight controller for three wave scenarios and two target damage values.\ For clarity, only every 100th value is displayed. Top: Accumulated damage over time. Middle: The selected weight index \iw over time. A lower index corresponds to a higher weighting for the damage cost. Bottom: Extracted energy over time. For the lower damage thresholds, the extracted energy reduces only by 0.09, 0.05, and \SI{0.06}{\mega\joule} for the three cases, respectively. \todo{bottom row can probably be removed and just described in the text.}}
	\label{fig:weight_control}
\end{figure*}

\subsection{Weight selection algorithm}
In this section, we evaluate the performance of the simple heuristic weight selection algorithm presented in section \ref{sec:weight_control} by simulating the system's behaviour for different sea states.\ 
We used a set of $\nw=15$ predetermined weights $w_2$, evenly distributed between 0.05 and 0.95, and chose $w_1$ such that $w_1+w_2=1$.\ 

For three different wave scenarios, we compare the performance of the \ac{mpc} with weight controller for two target damage values $\Jd = \left\{0.3, 0.5\right\}$.\ These threshold values (together with target time $\tbd=3000$ s) are used here for the sole purpose of exemplification, and they do not reflect real failure thresholds/failure time-scales (which are expectedly much larger). 
The performance of a weighting is evaluated every \SI{25}{\second}.\ 
An increase of the damage weighting is allowed after each evaluation, whereas a decrease is only conservatively allowed every two evaluations.

\begin{figure}[tb]
	\centering
	\fontsize{8}{0}\selectfont
	\def\svgwidth{0.49\textwidth}
	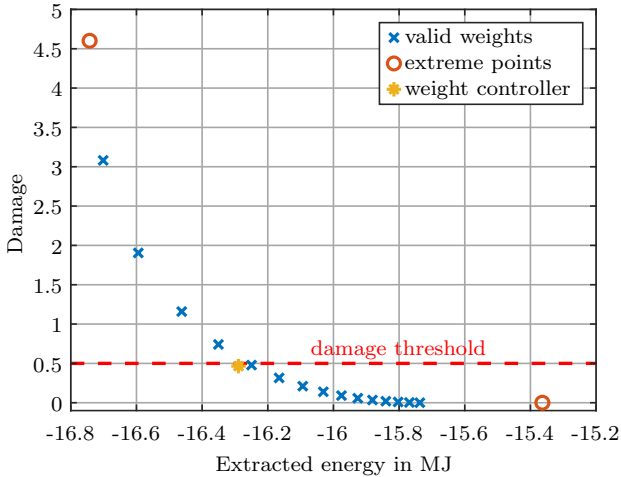
	\caption{The accumulated costs for the fixed weight MPC with the valid weights for the weight-controller and the extreme point approximations for the left case from \figref{fig:weight_control}.\ The costs resulting from the weight controller with threshold 0.5 dominate some of the fixed-weight costs.}
	\label{fig:mpc_pf}
\end{figure}

\figref{fig:weight_control} shows the performance of the heuristic weight control for  three different realisations of a same wave spectrum by displaying the accumulated damage in the top and the selected weighting index in the bottom plots.\ 
In the first example (left column), the excitation is composed by a sequence of small-amplitude oscillations (with nearly-constant amplitude and frequency). In this case, the damage weight is steadily decreased compared to the initial value until it reaches a steady value for the case with $\Jd=0.5$.\
The second example (central column) features a distribution of the weight index (center row) which leads to a smooth increase of the accumulated damage towards the target.\ 
Certain sea states might be characterised by phases during which the excitation $d$ has small amplitude. In these cases, the controller might select high values of $w_1$, which, in turn, might lead to sudden increases in damage as soon as the excitation amplitude increases again. This is the case for the results shown in the right column, where sudden increases in the damage are recorded.\ 
Even though the heuristic controller does not react fast enough, it only violates the respective threshold of 0.3 and 0.5 by less than 3 \% before selecting the weighting that accumulates the least damage. \\
Remarkably, the difference in extracted energy with the two threshold values for $\Jd$ in all three examples differs in small ranges: 0.55, 0.32, and 0.36 \%.\
To motivate this, we compare the weight-controlled \ac{mpc} with the fixed-weight MPC.\ 
In addition to the 15 weights used in the weight-controller, we also use $w_2=0.99$ and $w_2=0.01$ to approximate the extreme points that minimise one of the cost functions.\ 

\figref{fig:mpc_pf} shows that the difference in extractable energy for fixed weights is less than  \SI{6}{\percent}.\ 
The weight-controlled MPC yields more harvested energy than all fixed-weight MPCs that accumulate a damage $J_2\le\Jd$ over a time-frame of 3000 s. This is possible thanks to the fact that fixed-weight MPC solutions are not Pareto-optimal and they show deviations from the ground truth \ac{ocp} solution as 
shown in section \ref{sec:accuracy}.

	\section{Conclusion}
This work analyses the performance of model-predictive control (MPC) for dielectric elastomer generator-based wave energy converters under  stochastic waves excitation.\ 
Compared to previous work, an MPC approach is considered for control, as it allows  accounting for wave-by-wave changes in the excitation, while still accounting for energetic and damage cost functions.\ 
The MPC considerably deviates from ground truth solutions obtained by solving an optimum control problem (OCP) over an extremely long time frame, unless the prediction horizon covers a sufficiently large number of wave periods.\ 
As the sea state changes over time, the shape of the OCP Pareto front changes.\ 
As this makes achieving long-term objectives difficult, we propose a heuristic controller for selecting weight combinations that allow shifting the expected failure  of the generator towards a target time in the future.
	\bibliography{references.bib}
\end{document}